\documentclass[12pt,doublespace]{article}
\begin{document}
\begin{center}
GENERALIZED SYNCHRONIZATION IN LASER DEVICES WITH ELECTRO-OPTICAL FEEDBACK\\
E.M.Shahverdiev $^{1,2}$ and K.A.Shore $^{1}$\\
$^{1}$School of Electronic Engineering, Bangor University, Dean St.,Bangor, LL57 1UT, Wales, UK\\
$^{2}$Institute of Physics, H.Javid Avenue,33, Baku, AZ1143, Azerbaijan\\
~\\
ABSTRACT\\
\end{center}
We report on partial and global generalized synchronization in exemplar bi-directionally coupled non-identical chaotic systems with multiple time delays. We derive conditions for such synchronization regimes. The general approach is applied to the case of multiple time delay laser systems with electro-optical feedback. We also study the effect of signal sampling on the signatures of time delays in the autocorrelation function of the multiple time delay laser output.\\
\begin{center}
1.INTRODUCTION
\end{center}
\indent The synchronization of systems of coupled oscillators is important in many disciplines of science. In a chaos based secure communication scheme synchronization is vital for message decoding [1].\\
\indent Recently, delay differential equations (DDE) have attracted much attention in the field of nonlinear dynamics. The high complexity of the multiple time-delayed systems can provide a new architecture for enhancing message security in chaos based encryption systems [2]. In such communication systems message decoding would require chaos synchronization between multiple time-delayed transmitter and receiver systems.\\
\indent Several kind of synchronization can arise in interacting systems. Generalized synchronization, first studied for uni-directionally coupled systems, is defined as the presence of some functional relation between the states of response and drive, i.e. $y(t)=F(x(t))$ [3]. As underlined in [4], for secure chaos based communication schemes the use of generalized synchronization mode could be preferable in comparison with the non-generalized modes of synchronization. Recently, generalized synchronization between bi-directionally linearly coupled non-identical Lorenz and R{\"o}ssler models has been investigated numerically in [5] (by Zhigang {\it et al}), where the concept of generalized synchronization was extended to mutually coupled systems. It was shown that with increasing coupling strengths the system changes from partial to global generalized synchronization.\\
It is noted that bi-directionally coupled time delay systems with a single time delay were studied previously in by Heil {\it et al} [5]. Paper by T. Heil {\it et al} presents experimental and numerical investigations of the dynamics of two device-identical optically coupled semiconductor lasers exhibiting a delay in the coupling. They find well-defined time lag between the dynamics of the two lasers and an asymmetric physical role of the subsystems. Further, in paper by Fischer {\it et al} [5] the authors show that isochronous synchronization between delay  coupled oscillators can be achieved by relaying the dynamics via a third mediating element. In other words they consider three bi-directionally coupled oscillator in a line configuration and demonstrate that zero-lag synchronization can occur over long distances through relaying.\\
To the best of our knowledge partial and global generalized synchronization in time delayed systems has yet to be investigated.\\
\indent  This paper is devoted to the study of partial and global generalized synchronization between mutually coupled chaotic systems with multiple time-delays. We find analytically the stability condition of the global generalized synchronization mode between bi-directionally coupled multiple time-delayed systems and apply the approach to the multiple time delay semiconductor lasers with electro-optical feedback.\\
\indent In a time-delayed-chaos-based communication, identification of time delays can allow the eavesdropper to extract the message successfully using a simple local reconstruction of the time delay system. The delay time can be revealed by a number of means including via the autocorrelation coefficient [6]. With this in mind, in this paper we also investigate the effect of sampling rate on the autocorrelation coefficient of the laser system with electro-optical feedback. The principal result of this study is the demonstration that the recovery of the delay times from the autocorrelation coefficients of the laser output depends on the sampling rate of the signal.\\ 
\begin{center}
II.GENERAL APPROACH\\
\end{center}
\indent We consider synchronization between the bi-directionally coupled double-feedback systems of general form ,
\begin{equation}
\frac{dx}{dt}=-\alpha_{1} x + m_{1} f( x_{\tau_{1}})
+m_{2} f( x_{\tau_{2}}) + K_{1} (y-x),
\end{equation}
\begin{equation}
\frac{dy}{dt}=-\alpha_{2} y + m_{3} f( y_{\tau_{3}})
+ m_{4} f (y_{\tau_{4}}) + K_{2} (x-y),
\end{equation}
where $f$ is differentiable generic nonlinear function; $\alpha_{1,2}$ are the relaxation
coefficients for dynamical variables $x$ and $y$; $m_{1,2}$ and $m_{3,4}$ are the
feedback rates for systems $x$ and $y$, respectively; $\tau_{1,2}$ and $\tau_{3,4}$
are the feedback delay times in the coupled systems; $K_{1,2}$ are the coupling rates between  $x$ and $y.$
Throughout this paper $x_{\tau}\equiv x(t-\tau)$.\\
We find that under the conditions
\begin{equation}
\alpha_{1}=\alpha_{2},m_{1}=m_{3}, m_{2}=m_{4},K_{1}=K_{2}=K,\tau_{1}=\tau_{3},\tau_{2}=\tau_{4}
\end{equation}
Eqs.(1) and (2) admit the synchronization manifold
\begin{equation}
y=x.
\end{equation}
This follows from the dynamics of the error $\Delta=x-y$ (with conditions (3))
\begin{equation}
\frac{d\Delta}{dt}= -\alpha_{1}\Delta + m_{1} \Delta_{\tau_{1}}f'(x_{\tau_{1}})
+m_{2}\Delta_{\tau_{2}}f'(x_{\tau_{2}})-2K\Delta.
\end{equation}
obtained by use of a Taylor series. Here $f'(x)$ denotes for derivative of $f$ and the derivative
should be bounded. Solution $y=x$ under the conditions (3) is also follows from the symmetry of Eqs.(1) and (2). 
According to the Lyapunov-Razumikhin functional approach (see, e.g. [2] and references there-in), the sufficient stability 
condition for the synchronization regime $y=x $ (4) as:
\begin{equation}
\alpha_{1} +2K > \vert m_{1}(\sup f'( x_{\tau_{1}}))\vert +\vert m_{2}(\sup f'( x_{\tau_{2}}))\vert.
\end{equation}
Here $\sup f'(x)$ stands for the supremum of $f'$ with respect to the appropriate norm.\\
\indent Next we consider the effect of parameter mismatches on
synchronization . For the parameter mismatches,
$\alpha_{1}\neq\alpha_{2},m_{1}\neq m_{3}, m_{2}\neq
m_{4},K_{1}\neq K_{2},\tau_{1}\neq \tau_{3},\tau_{2}\neq \tau_{4}$
is clear that complete synchronization,Eq.(4) is no longer the
synchronization regime for coupled systems (1) and (2). Then for
such a case one can try using the auxiliary system method to detect
generalized synchronization between bi-directionally coupled $x$
and $y$ [5]. According to [5] bi-directionally coupled systems
$x$ and $y$ can be considered as {\it two} uni-directionally
coupled systems driven by $x$ and $y$. Then for each of the
uni-directionally coupled systems one can apply the concept of
generalized synchronization: that is given another identical driven
auxiliary system $z(t)$, generalized synchronization between
$x(t)$ and $y(t)$ is established with the achievement of complete
synchronization between $y(t)$ and $z(t)$. Thus, in the case of
bi-directionally coupled systems for generalized synchronization
one has to investigate the possibility of two complete
synchronization manifolds, one for $x$ driving, and another one for $y$
driving. Depending on the achievement of complete synchronization due to the coupling
strengths between the systems, partial (when the complete
synchronization occurs between the systems driven either by $x$ or
$y$) and global (when complete synchronization occur between the
systems driven by both $x$ and $y$) generalized synchronization
can arise. Applying the error dynamics approach to investigate the complete
synchronization between the systems (driven by $y$)
\begin{equation}
\frac{dx}{dt}=-\alpha_{1} x + m_{1} f( x_{\tau_{1}})
+m_{2} f( x_{\tau_{2}}) + K_{1}(y-x),
\end{equation}
\begin{equation}
\frac{dx_{1}}{dt}=-\alpha_{1} x_{1} + m_{1} f( x_{1,\tau_{1}})
+m_{2} f( x_{1,\tau_{2}}) + K_{1}(y-x_{1}),
\end{equation}
and the systems (driven by $x$)
\begin{equation}
\frac{dy}{dt}=-\alpha_{2} y + m_{3} f( y_{\tau_{3}})
+ m_{4} f (y_{\tau_{4}}) + K_{2} (x-y),
\end{equation}
\begin{equation}
\frac{dy_{1}}{dt}=-\alpha_{2} y_{1} + m_{3} f( y_{1,\tau_{3}}) +
m_{4} f (y_{1,\tau_{4}}) + K_{2}(x-y_{1}),
\end{equation}
we find the sufficient stability conditions for the complete
synchronization $x=x_{1}$
\begin{equation}
K_{1} > \vert m_{1}(\sup f'( x_{\tau_{1}}))\vert +\vert m_{2}(\sup f'( x_{\tau_{2}}))\vert -\alpha_{1}.
\end{equation}
and for $y=y_{1}$
\begin{equation}
K_{2} > \vert m_{3}(\sup f'( y_{\tau_{3}}))\vert +\vert m_{4}(\sup f'( y_{\tau_{4}}))\vert -\alpha_{2}.
\end{equation}
Then for $(K_{1},K_{2}) \geq \max(K_{1},K_{2})$ complete synchronization occurs, which means that we have a
case of global generalized synchronization between the bi-directionally coupled systems $x$ and $y.$\\
We notice that the stability conditions (6), (11) and (12)
derived using the Lyapunov-Razumikhin approach is a sufficient one: it assures a high quality
synchronization for a coupling strength estimated from the stability condition, but does
not forbid the possibility of synchronization with smaller coupling strengths. The threshold
coupling strength can be estimated by the dependence of the maximal conditional Lyapunov exponent
$\lambda$ of the error dynamics on coupling strength K:i.e. from $\lambda (K)=0$ [7]. Suppose that
$K_{1}^{n}$ and $K_{2}^{n}$ are the value of coupling strengths found from $\lambda (K)=0$ for
auxiliary systems (7-8) and (9-10), respectively. Then for $(K_{1},K_{2}) < \min(K_{1}^{n},K_{2}^{n})$ there is no
synchronized state, for $\min (K_{1}^{n},K_{2}^{n})\leq (K_{1},K_{2}) \leq \max(K_{1}^{n},K_{2}^{n})$ there is partial and
for $(K_{1},K_{2}) \geq \max(K_{1}^{n},K_{2}^{n})$ there is global generalized synchronization.\\
\begin{center}
III.WAVELENGTH CHAOS MODEL
\end{center}
\indent Synchronization of chaotic semiconductor lasers is of
significant practical importance, as these lasers have potential
application in high-speed secure communications [1]. The laser
system considered in this paper is an electrically tunable
Distributed Bragg Reflector (DBR) laser diode with electro-optical
feedback. This system was proposed in [8] as a chaotic wavelength signal generator for chaos based secure communication.
The wavelength of the chaotic carrier is described by the
following dynamical equation:
\begin{equation}
T\frac{\lambda(t)}{dt}=-\lambda (t) + \beta_{\lambda}\sin^{2}(\frac{D\pi}{\Lambda^{2}_{0}}\lambda (t-\tau) -\Phi_{0})
\end{equation}
where $\lambda$ is the wavelength deviation from the center wavelength $\Lambda_{0}$;$D$ is the optical path difference
of the birefringent plate that constitutes the nonlinearity;$\Phi_{0}$ is the feedback phase;$\tau$-the feedback loop
delay time; $T$ is the time response in the feedback loop;$\beta_{\lambda}$ is the feedback strength.
With $x=\frac{\pi D\lambda}{\Lambda^{2}_{0}}$ and $m=\frac{\pi D\beta_{\lambda}}{\Lambda^{2}_{0}}$ we
rewrite Eq.(13) in the following normalized  form:
\begin{equation}
\frac{dx(t)}{dt}=-\alpha x(t) + m \sin^{2}(x_{\tau} -\Phi_{0}).
\end{equation}
Notice that in (14) we have scaled the time with $\alpha T$($\alpha$ is the relaxation coefficient). In the following we consider chaos synchronization between
Bi-directionally linearly coupled electro-optical semiconductor lasers with double feedbacks.\\
\begin{center}
IV.BIDIRECTIONALLY COUPLED SYSTEMS
\end{center}
The lasers to be synchronized are described by the following equations
\begin{equation}
\frac{dx(t)}{dt}=-\alpha x(t) + m_{1} \sin^{2}(x_{\tau_{1}} -\Phi_{0}) + m_{2} \sin^{2}(x_{\tau_{2}} -\Phi_{0}) + K_{1}(y-x),
\end{equation}
\begin{equation}
\frac{dy(t)}{dt}=-\alpha y(t) + m_{3} \sin^{2}(y_{\tau_{3}} -\Phi_{0}) + m_{4} \sin^{2}(y_{\tau_{4}} -\Phi_{0}) + K_{2}(x-y),
\end{equation}
Applying the general approach from Section II we find that systems (15) and (16) can be synchronized on
the synchronization manifold (4) under the existence conditions (3).
The sufficient stability condition for the synchronization manifold $y=x$ for the coupled systems (15-16) is :
\begin{equation}
2K > (m_{1} +m_{2})- \alpha.
\end{equation}
\indent Considering the effect of parameter mismatches on the
synchronization regime by use of the results of Section II, we
establish that the global generalized synchronization between $x$
and $y$ occurs if the coupling strengths $(K_{1},K_{2})\geq\max(K_{1}, K_{2})$ where $K_{1}$ and $K_{2}$
satisfy the inequalities:
\begin{equation}
K_{1} > (m_{1} +m_{2})- \alpha,
\end{equation}
\begin{equation}
K_{2} > (m_{3} + m_{4})- \alpha .
\end{equation}
Under the condition (18) complete synchronization occurs between the $y$ driven auxiliary systems and (19) is the sufficient stability
condition for the complete synchronization between the auxiliary systems driven by $x$.\\
\begin{center}
V.NUMERICAL SIMULATIONS AND DISCUSSIONS
\end{center}
\indent Numerical simulations fully support the analytical results. In the simulations significant attention is given to the study of partial and global generalized synchronizations between non-identical systems. This is due to the possible advantages of generalized synchronizations in secure chaos based communication schemes, as underlined in [4]. With the security issue in mind, we also investigate the effect of the sampling rate on the autocorrelation function modulations.\\
Numerical modelling of Eqs.(15) and (16) were conducted using DDE23 program in Matlab (R2008b). Below we present results of numerical simulations for symmetrical coupling ($K=K_{1}=K_{2}$) to emphasize the fact that, despite the
symmetry in the coupling in bi-directionally coupled systems there is asymmetry in achieving synchronization between these directions.\\
We have also calculated the cross-correlation coefficients [9] using the formula
\begin{equation}
C(\Delta t)= \frac{<(x(t) - <x>)(y(t+\Delta t) - <y>)>}{\sqrt{<(x(t) - <x>)^2><(y(t+ \Delta t) - <y>)^2>}},
\end{equation}
where $x$ and $y$ are the outputs of the lasers, respectively; the brackets$<.>$
represent the time average; $\Delta t$ is a time shift between laser outputs. This coefficient
indicates the quality of synchronization: C=1 means perfect synchronization.\\
\indent Figure 1 shows complete synchronization between mutually coupled identical laser systems with double time delays.\\ 
Before considering synchronization between non-identical systems, we would like to dwell on the one aspect of the security for chaos based communication schemes, namely the effect of the signal's sampling rate on the autocorrelation function.
Figure 2 demonstrates the autocorrelation coefficient (here the autocorrelation coefficient will be denoted by $C_{A}$ and obtained from equation (20) when $x=y$ ) for the output of $x$ laser, Eqs.(1-2) for double time delays, with $\tau_{1}=3, \tau_{2}=5$, $m_{1}=2.3, m_{2}=2.5,\Phi_{0}=\pi/4, \alpha =1.$ It is clearly seen that time delays can be easily recovered from the autocorrelation coefficient, as it exhibits extrema at values of  
the time delays or their multiples and combinations. In this connection we would like to emphasize the following important point. The number of data points i.e. sampling rate  used to calculate autocorrelation coefficient is highly significant to detect the modulations of the autocorrelation coefficient at time delays. In other words the sampling rate is of importance for the security of chaos based communication systems.
Indeed for figure 2 the number of data points used was 65, which allowed recovery of the time delays. In figure 3 the number of data points was 25. Figure 4 contains 5000 data points. It is noted that both undersampling (fig.3) and oversampling (fig.4) do not allow extraction  of the correct time delays from the modulations of the autocorrelation coefficients.\\
\indent In the remainder of the paper we focus on the synchronization between non-identical systems with double time delays. In figures 5-7 we present the results 
of numerical simulations of Eqs.(15-16) for 
for $\Phi_{0}=\pi/4$,$m_{1}=20, m_{2}=15, m_{3}=4, m_{4}=3,\tau_{1}=2, \tau_{2}=1,\tau_{3}=6,\tau_{4}=7.$
Figure 5 demonstrates the unsynchronized state between non-identical lasers for coupling strengths $K_{1}=K_{2}=0.1$. It is noted that for such coupling strengths the error signal between the auxiliary systems in both directions does not approach zero. In figure 6 partial generalized synchronization between the interacting lasers is presented for the coupling strengths $K_{1}=K_{2}=3.2.$ In the case of partial generalized synchronization the error signal in one direction ($y_{1}-y$) approaches zero with time, i.e. synchronization is achieved between the auxiliary systems driven by laser $x$. In the reverse direction there is no synchronization,i.e. the synchronization error signal $x_{1}-x$ does not tend to zero with time.\\
\indent Figure 7 presents the case of global generalized synchronization between mutually coupled lasers $x$ and $y$ for $K_{1}=K_{2}=12.$ In this case the auxiliary systems driven by both lasers are synchronized, i.e. both error signals $y_{1}-y$ and $x_{1}-x $ tend to zero. We notice that in the case of partial generalized synchronization, despite the symmetry in the coupling, the coupling directions are not equivalent; there is synchronization in one direction and no synchronization in the reverse direction. In other words there is  always a master laser and a slave laser, despite the symmetrical
coupling. With increasing coupling strengths this asymmetry in the status of lasers is eliminated: both systems $x$ and $y$ control each other in the sense that both driven systems (by $x$ and $y$) are synchronized and eventually $x$ and $y$ enter a state of global generalized synchronization.\\
It is noted that the range of coupling strengths for the asymmetry is determined by the values of $K$ when the maximum conditional Lyapunov exponents for the auxiliary systems reach zero. Figure 8 depicts the maximum transversal (conditional)Lyapunov exponents $\lambda_{x}^{T}$($\triangle $)
 and $\lambda_{y}^{T}$ ($\diamondsuit $) versus the coupling strength for the auxiliary systems $x_{1}$ and $x$ driven by $y$ and $y_{1}$ and $y$ systems driven by $x,$ respectively. As evidenced from the numerical simulations, for the data used in our simulations the value of the maximum transversal (conditional) Lyapunov exponent reaches zero $\lambda^{T}_{y}\approx 0$ at $K_{1}^{n}\approx 3.1$(when the auxiliary systems $y_{1}$ and $y$ driven by $x$ laser is synchronized) and $\lambda^{T}_{x}\approx 0$ at $K_{2}^{n} \approx 9.2$(when the auxiliary systems $x_{1}$ and $x$ driven by $y$ laser are synchronized). As suggested in [5] the more chaotic system plays the role of a master, and less chaotic systems assume the role of the slave. The simulation results thus support the fact that $x$ laser is more chaotic, and hence it behaves as the master laser, as the largest Lyapunov exponent for $x$ laser is $\lambda_{x max}\approx 0.93$, and for $y$ laser $\lambda_{y max} \approx 0.10.$ It is noted that x laser is more chaotic than y laser 
because of higher feedback values.\\
\indent Finally we consider the role of parameter mismatches to achieve synchronization between uni-directionally linearly coupled 
non-identical time-delayed systems and show that parameter mismatches are of crucial importance to achieve synchronization. 
\begin{equation}
\frac{dx(t)}{dt}=-\alpha_{1}x(t) + m_{1} \sin^{2}(x_{\tau_{1}} -\Phi_{0}) + m_{2} \sin^{2}(x_{\tau_{2}} -\Phi_{0}),
\end{equation}
\begin{equation}
\frac{dy(t)}{dt}=-\alpha_{2} y(t) + m_{3} \sin^{2}(y_{\tau_{3}} -\Phi_{0}) + m_{4} \sin^{2}(y_{\tau_{4}} -\Phi_{0}) + Kx_{\tau_{3}},
\end{equation}
Let there be certain mismatches between the relaxation coefficients:$\alpha_{1}=\alpha -\delta,$ and $\alpha_{2}=\alpha + \delta.$
Then by investigating the dynamics of error $x_{\tau_{3}}-y=\Delta $ it is straightforward to establish that synchronization  $y=x_{\tau_{3}}$ is possible 
under existence conditions $m_{1}=m_{3},m_{2}=m_{4},\alpha_{2}-\alpha_{1}=K$ and is stable if $\alpha_{2}>\vert m_{3}\vert + \vert m_{4}\vert.$\\
Detailed study establishes that independent of the relation between the delay times in the coupled systems and the coupling delay time, only retarded synchronization with the coupling delay time is obtained (figure 9). Most importantly with or without parameter mismatch neither complete nor anticipating synchronization occurs.\\
We mention that, for example in the case of non-linear (sinusoidal) coupling for identical drive and response systems with electro-optical feedbacks, depending on the relation between the feedback delays and the coupling delay time retarded, complete or anticipating synchronization can occur [10-11]. These results are of significant interest in the context of relationship between parameter mismatches, coupling forms and synchronization. Indeed, having in mind possible practical applications of anticipating chaos synchronization in secure communications, in the control of delay-induced instabilities in a wide range of non-linear systems, etc, by choosing the "appropriate" parameter mismatches and coupling forms certain types of synchronization can be achieved.\\
\begin{center}
VI.CONCLUSIONS\\
\end{center}
\indent We have investigated partial and global generalized synchronization regimes between two bi-directionally coupled time-delayed systems with
multiple delays. We have derived the conditions for such synchronization regimes. It is well known that in chaos based communications chaos synchronization between the transmitter and receiver is vital for the message decoding. As such synchronization conditions are of importance for communication between the systems. We have tested in the case of laser systems with electro-optical feedback. The results are important from the point of view of transition from the disordered state of interacting complex systems to the partial and global order states. These findings are also of importance for secure chaos-based communication systems in the context of advantages of generalized synchronization mode in comparison with the non-generalized regimes of synchronization.\\
Also by studying the effect of the sampling rate on the signatures of time delays in the autocorrelation function of the multiple time delay laser output we have demonstrated that the optimal sampling rate is essential to recover the time delays. This property is also important for secure chaos-based communication systems.\\
~\\
{\it Acknowledgements}.-This research was supported by a Marie Curie Action  within the $6^{th}$ European Community Framework Programme Contract.\\
\newpage
\begin{center}
Figure captions
\end{center}
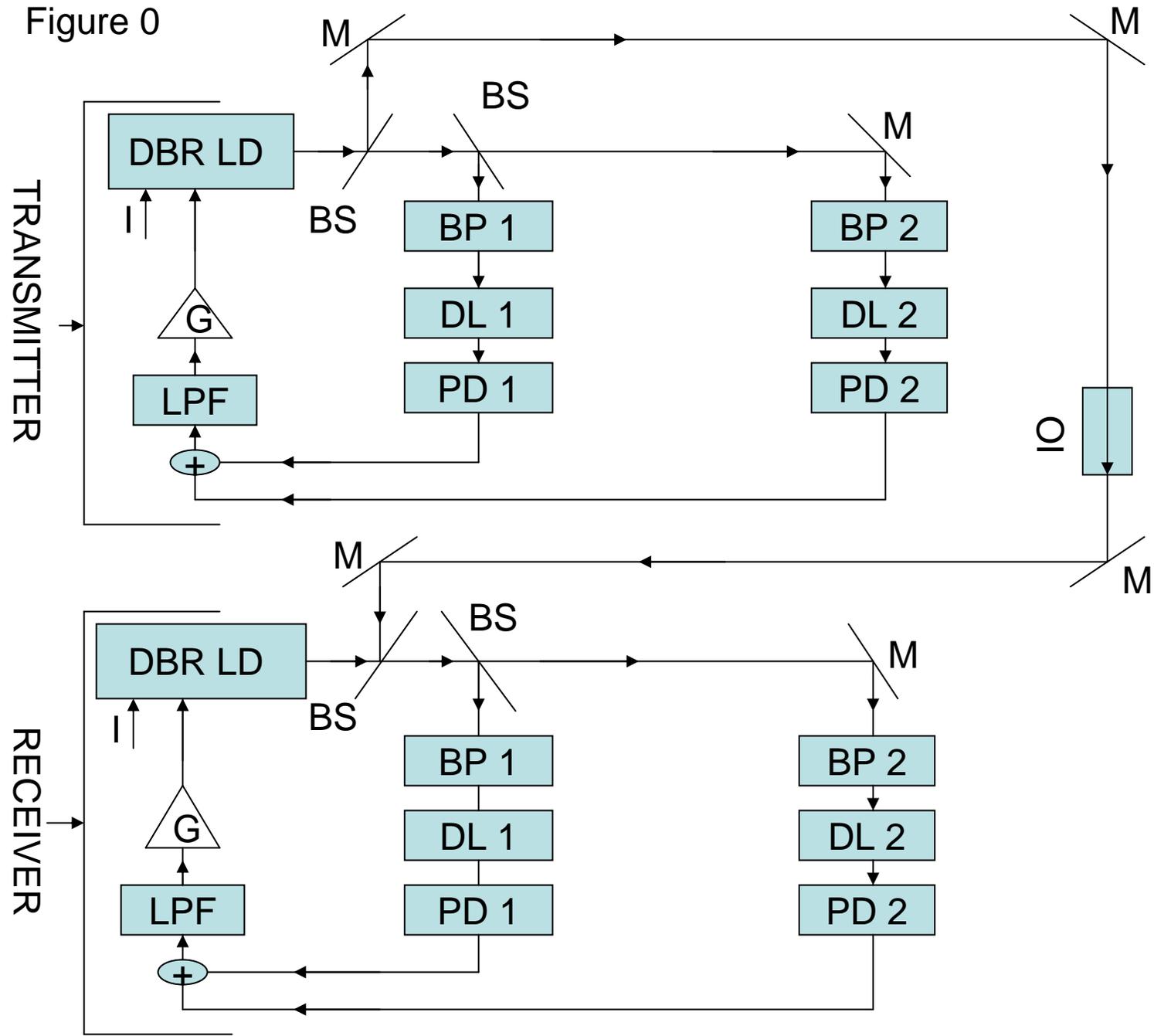
\noindent FIG.0.Schematic experimental set-up for wavelength chaos synchronization between the transmitter and receiver laser diodes with electooptical feedback: DBR LD1 and DBR LD2 are the Distributed Bragg Reflector transmitter and receiver laser diodes, respectively; BS, beamsplitter; M, mirror; BP, birefringent plate between crossed polarizers(not shown in the figure); PD, photodiode; DL, delay line; LPF, low-pass filter; G, optoelectronic gain; 
I, DBR-section injection current; OI, optical isolator to provide unidirectional coupling between laser diodes.\\
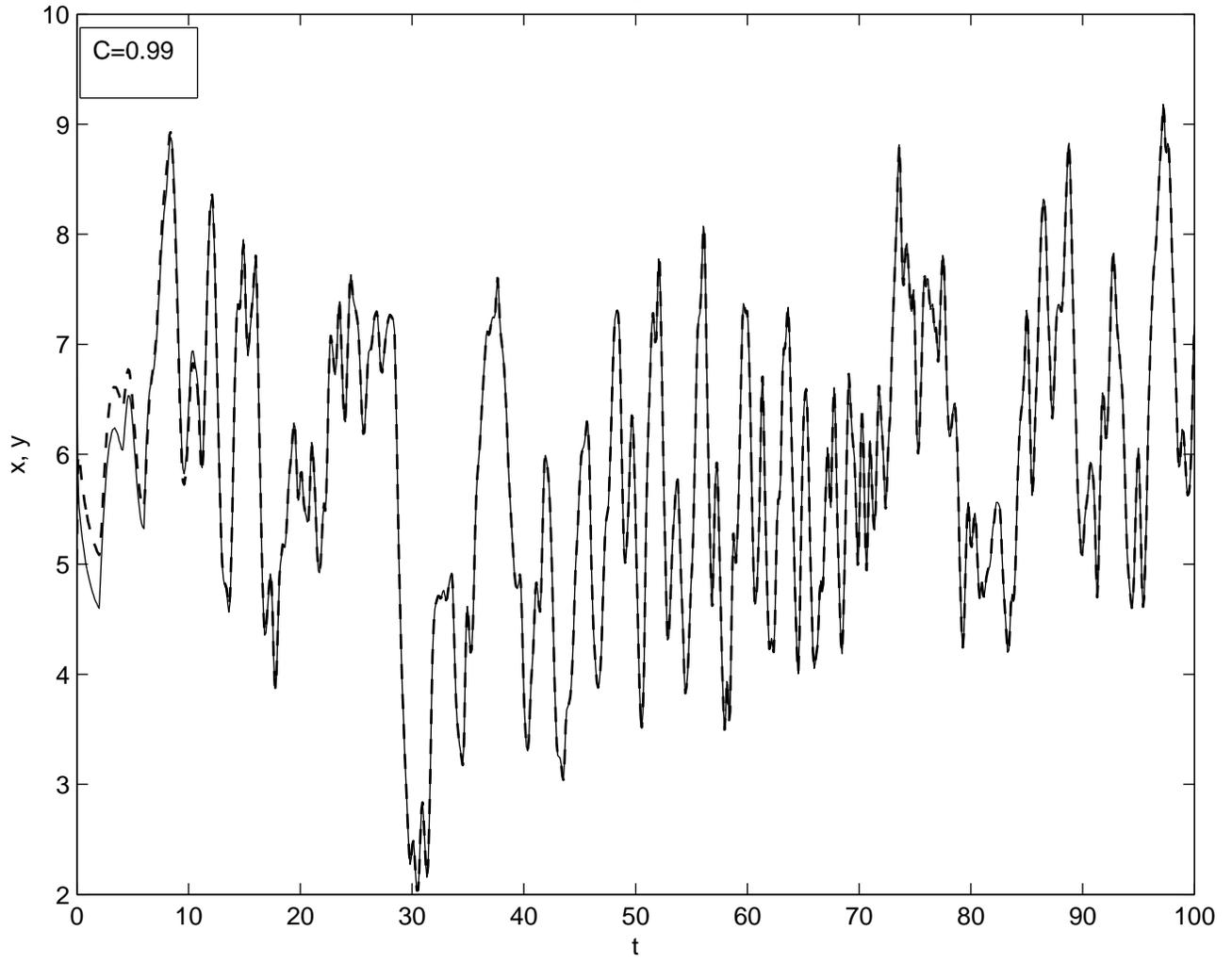
\noindent FIG.1.Numerical simulation of mutually coupled identical electro-optical laser systems, Eqs.(15-16). Complete synchronization: time series of $x(t)$ (solid line) and $y(t)$(dashed line) for $\alpha =1,\Phi_{0}=\pi/4$,$m_{1}=m_{3}=5, m_{2}=m_{4}=7,K_{1}=K_{2}=9,\tau_{1}=\tau_{3}=2, \tau_{2}=\tau_{4}=6;$C is the correlation coefficient between $x$ and $y.$ Dimensionless units.\\
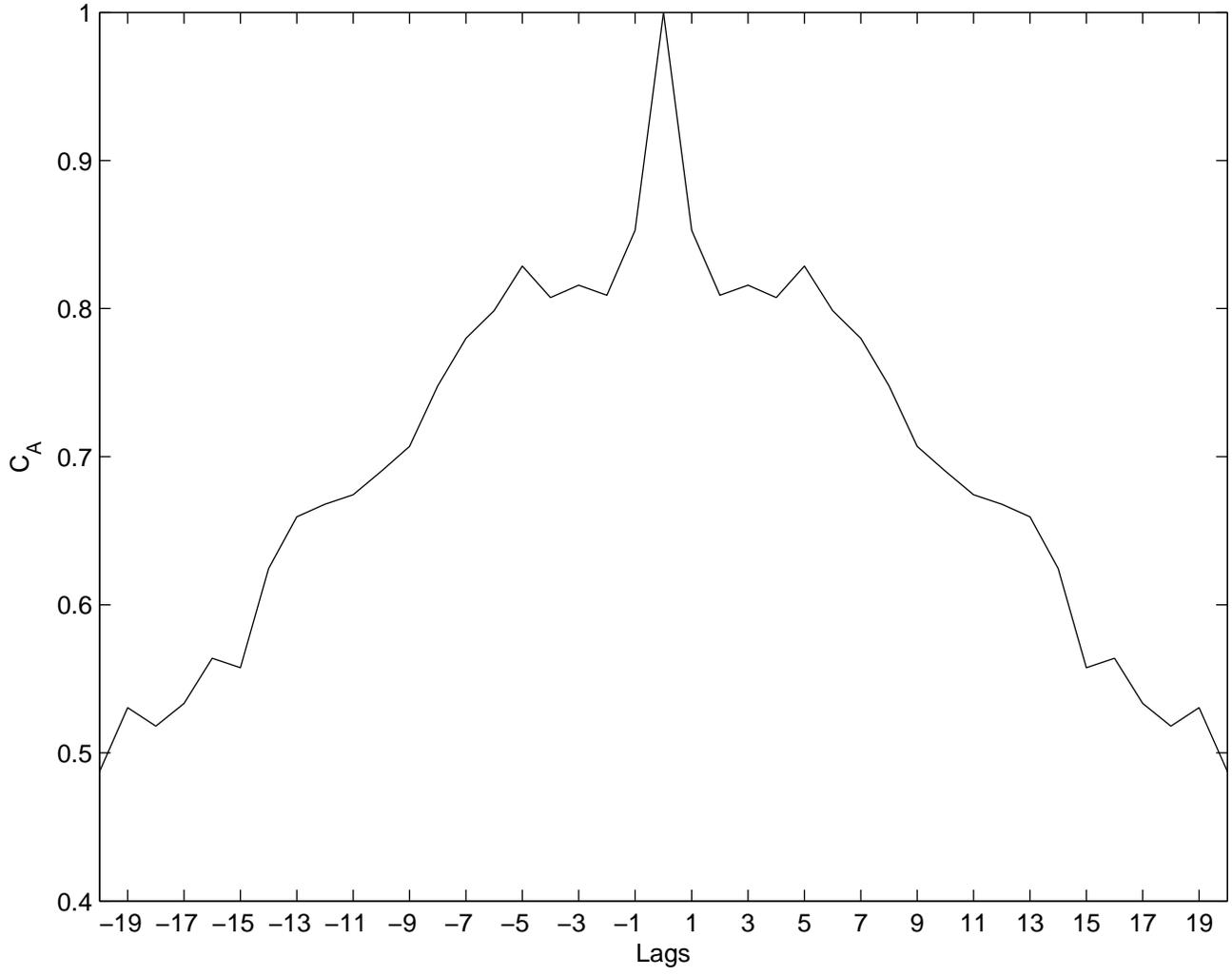
\noindent FIG.2. The autocorrelation coefficient $C_{A}$ of $x$ laser output for parameters, Eqs.(15)($K_{1}=0$) for $\tau_{1}=3, \tau_{2}=5,  m_{1}=5, m_{2}=7,\alpha=2,\Phi_{0}=\pi/4.$ The number of data points N=65; Sampling time=300/N. Dimensionless units.\\
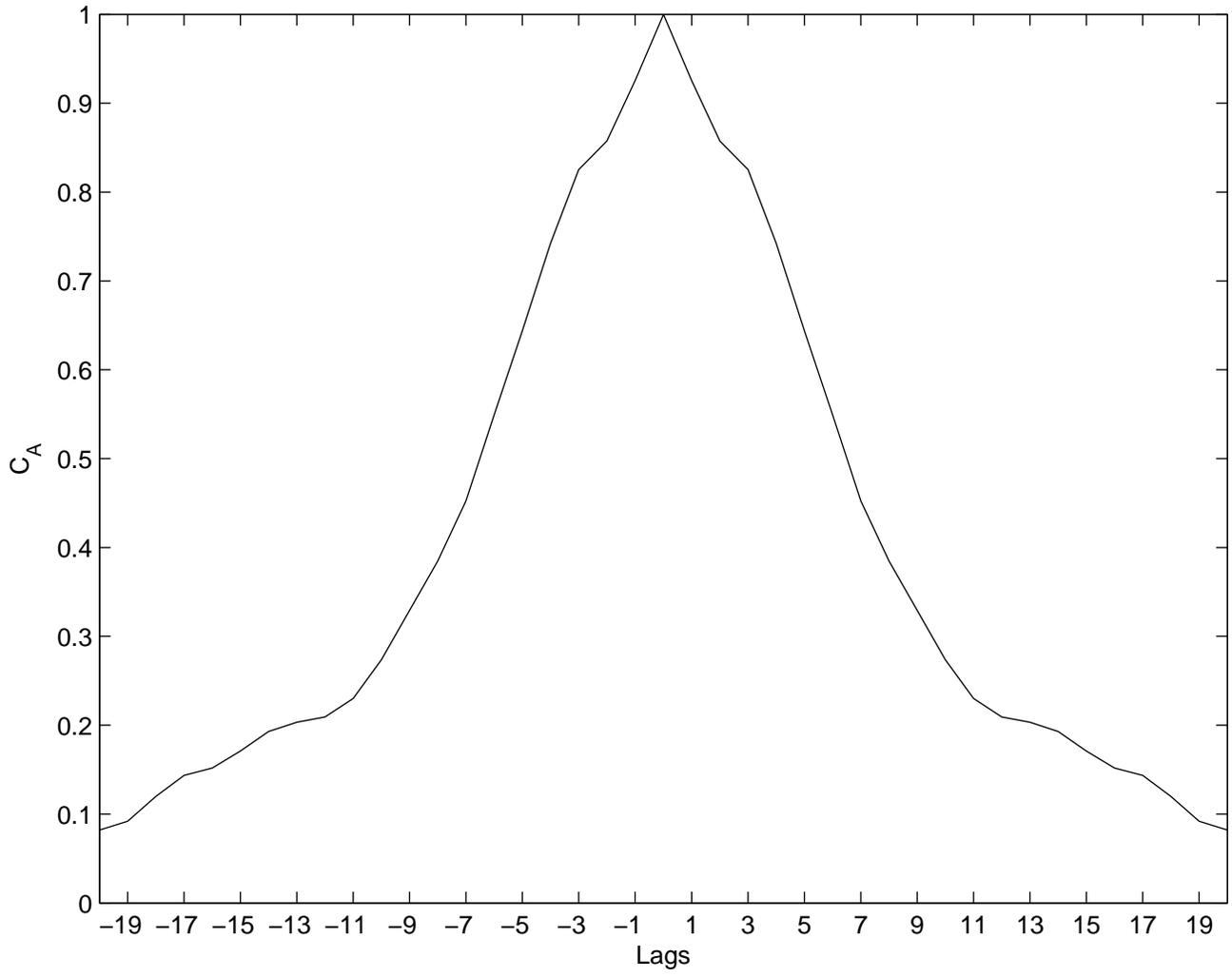
\noindent FIG.3. The autocorrelation coefficient $C_{A}$ of $x$ laser output for parameters, Eqs.(15)($K_{1}=0$) for $\tau_{1}=3, \tau_{2}=5,  m_{1}=5, m_{2}=7,\alpha=2,\Phi_{0}=\pi/4.$  The number of data points N=25; Sampling time=300/N.Dimensionless units.\\
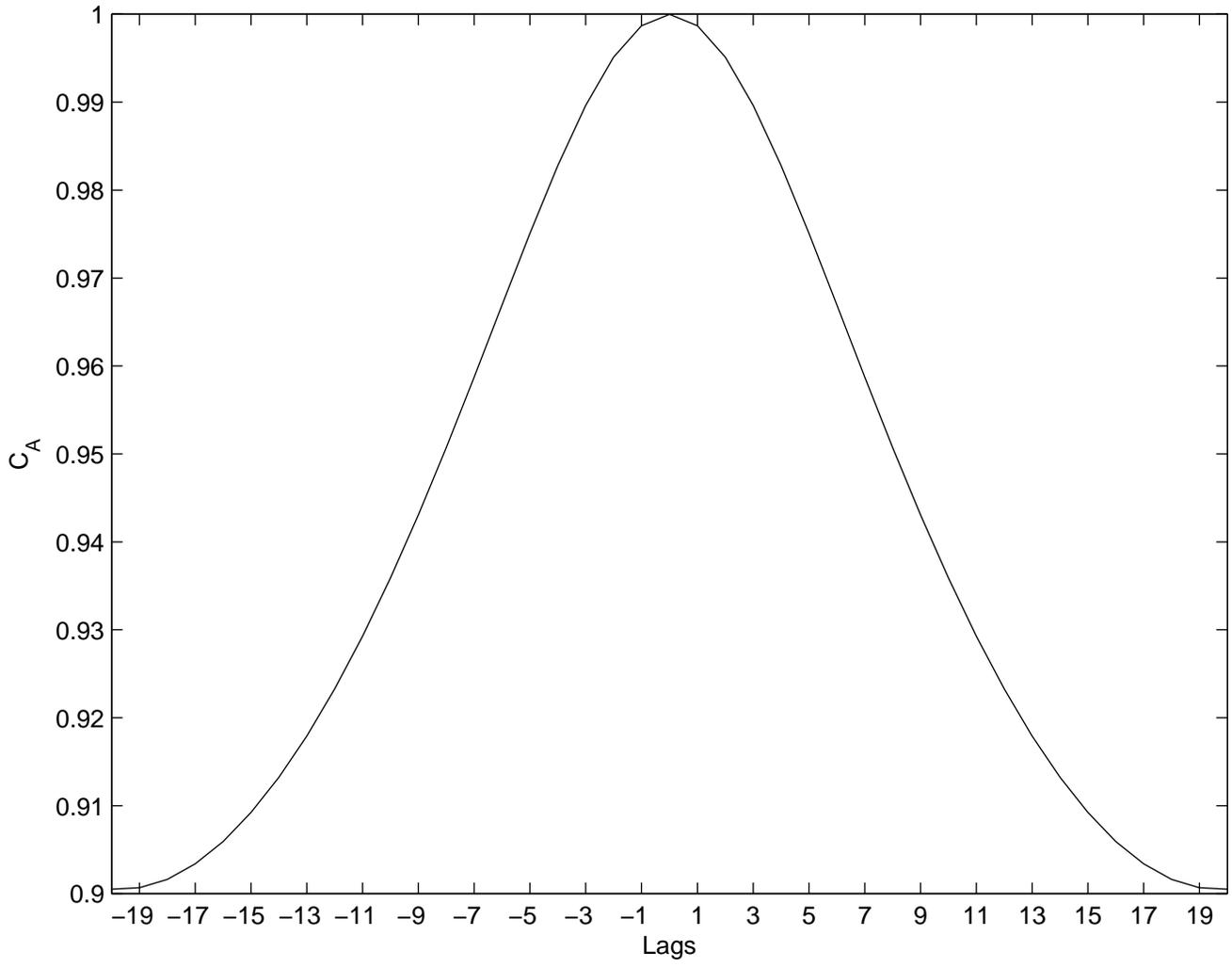
\noindent FIG.4. The autocorrelation coefficient $C_{A}$ of $x$ laser output for parameters, Eqs.(15)($K_{1}=0$) for $\tau_{1}=3, \tau_{2}=5,  m_{1}=5, m_{2}=7,\alpha=2,\Phi_{0}=\pi/4.$  The number of data points N=5000; Sampling time=300/N. Dimensionless units.\\
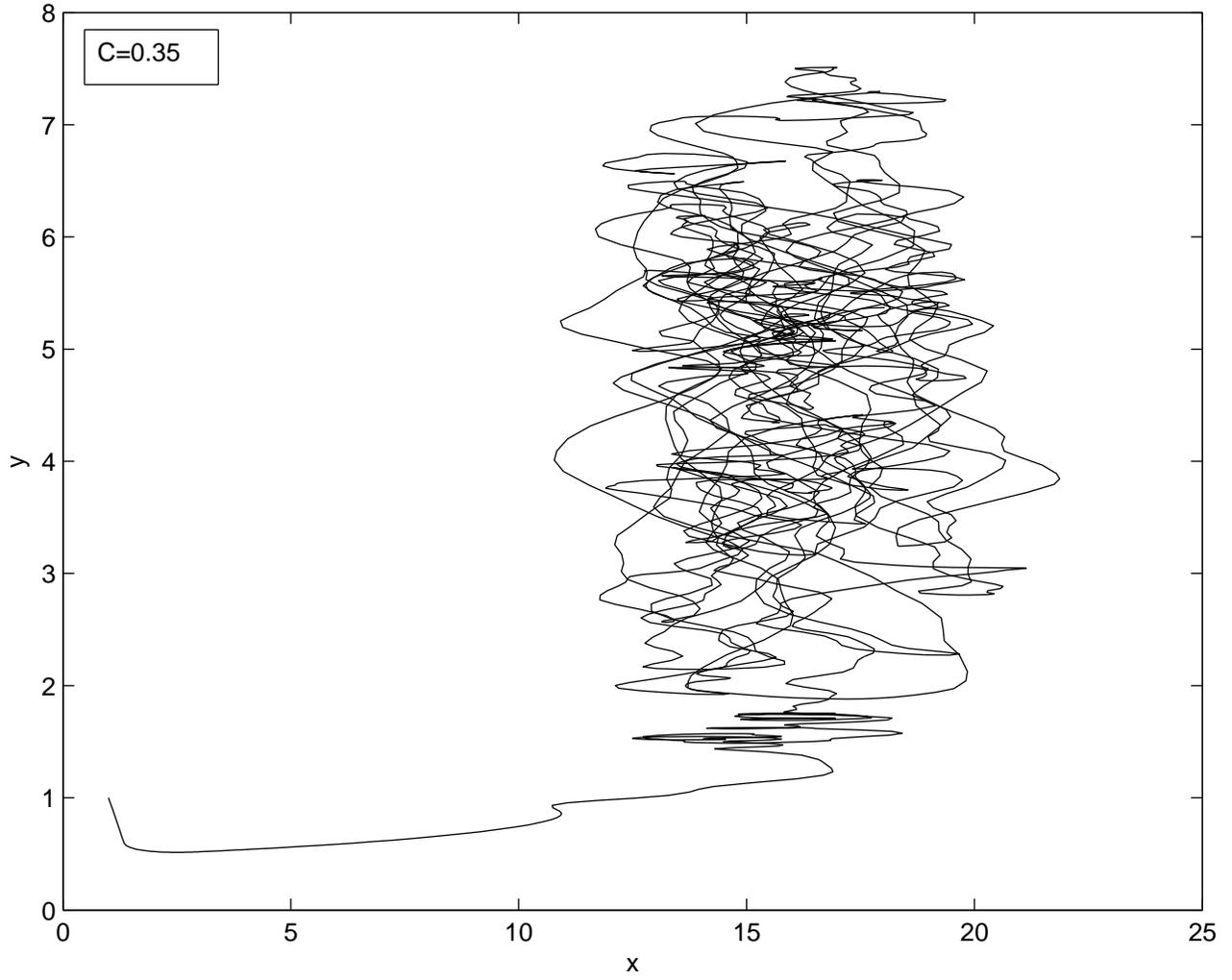
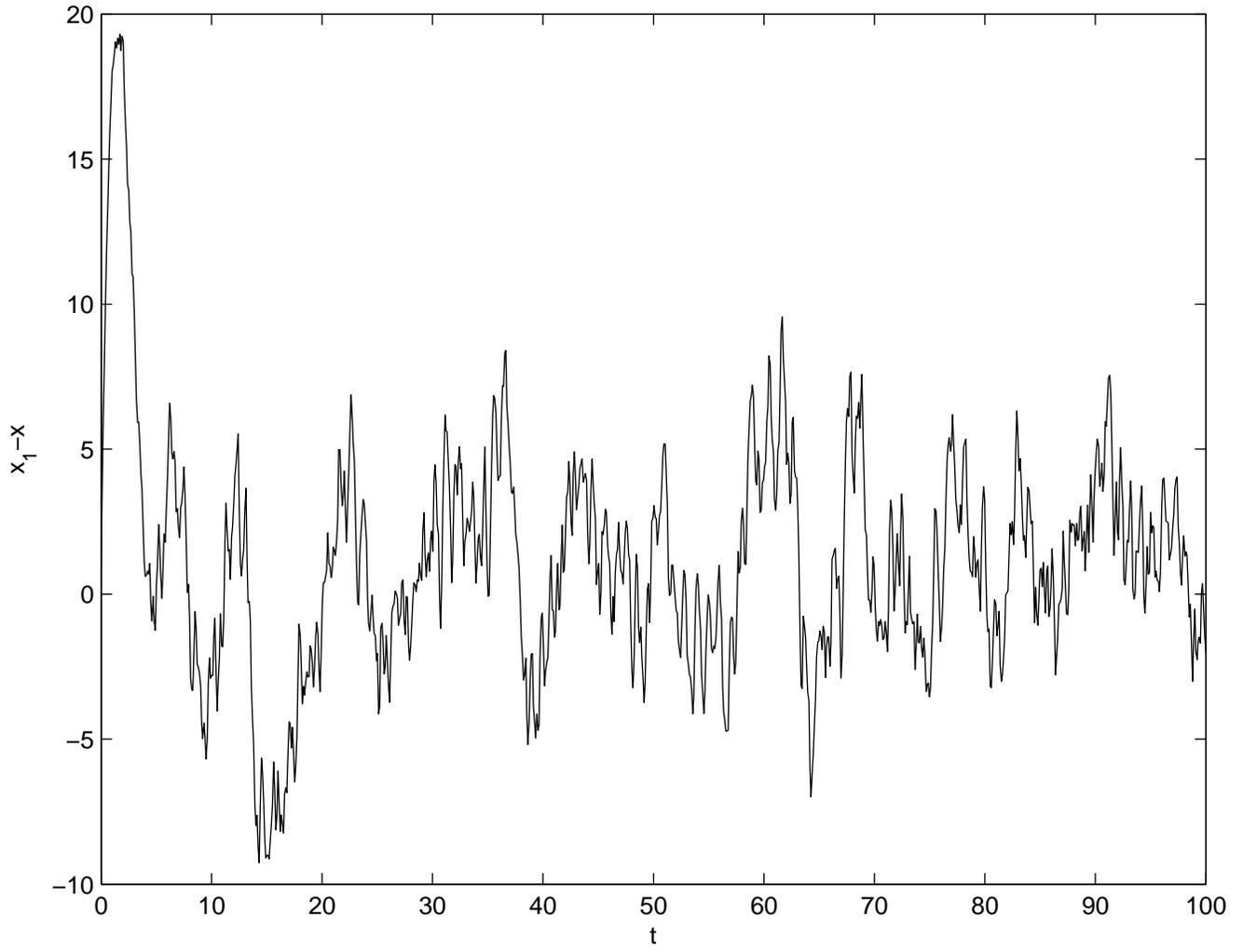
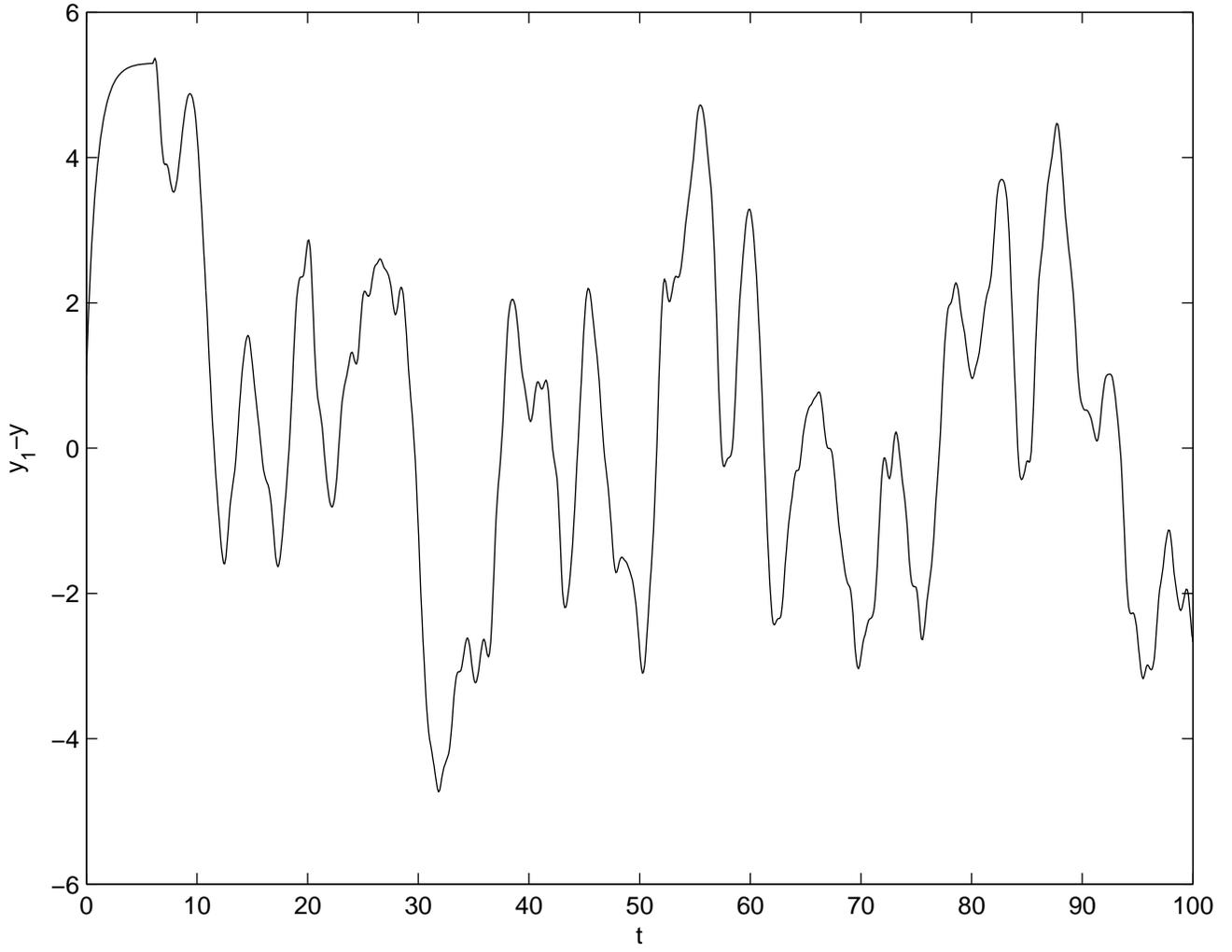
\noindent FIG.5. Numerical simulation of mutually coupled non-identical electro-optical laser systems, Eqs.(15-16): Unsynchronized state between $x$ and $y$ for $\Phi_{0}=\pi/4$,$m_{1}=20, m_{2}=15, m_{3}=4, m_{4}=3,K_{1}=K_{2}=0.1,\tau_{1}=2, \tau_{2}=1,\tau_{3}=6,\tau_{4}=7;$(a)correlation plot between $x$ and $y$ laser outputs;(b)error $x_{1}-x$ dynamics;(c)error $y_{1}-y$ dynamics.C is the correlation coefficient between $x$ and $y.$ Dimensionless units.\\
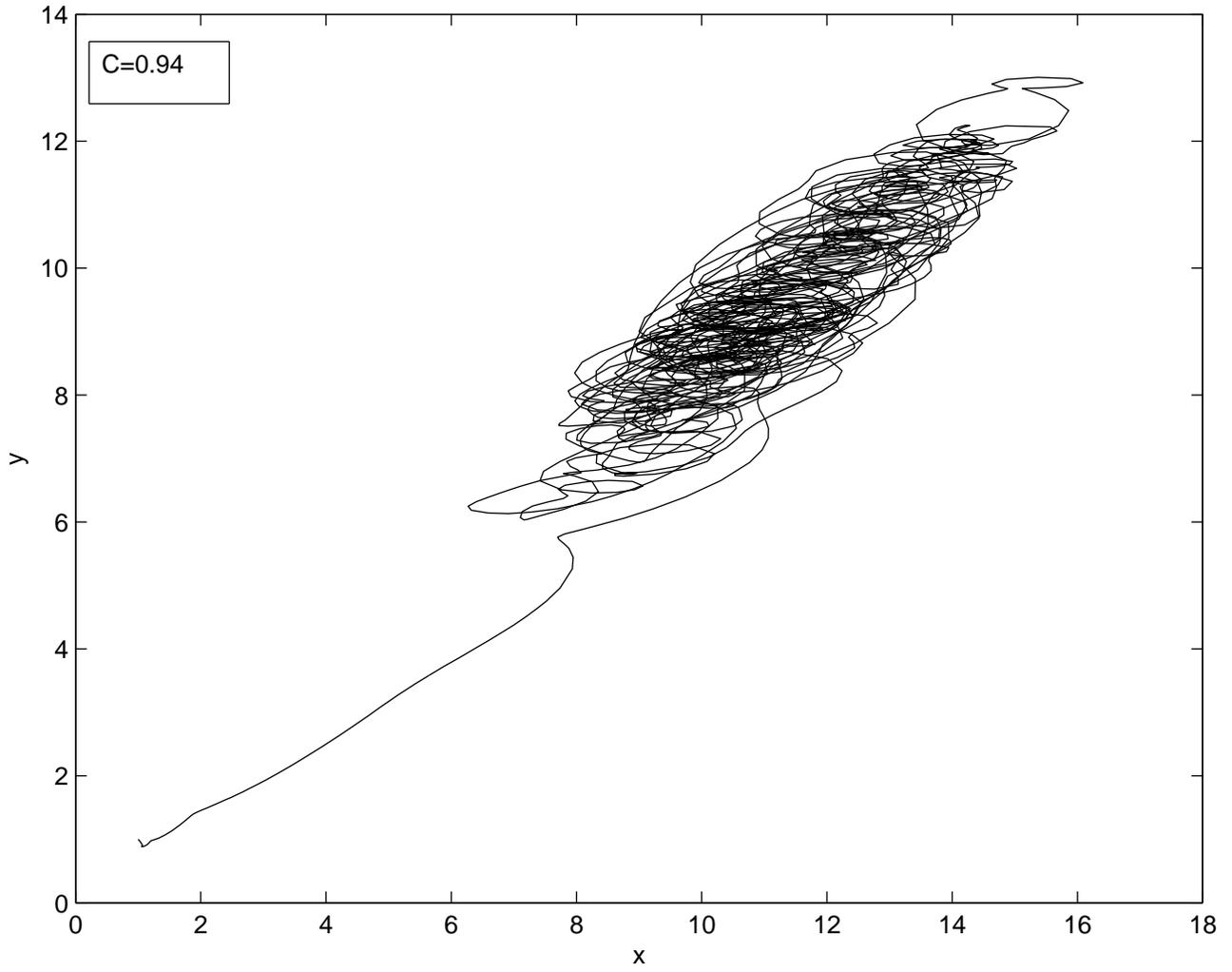
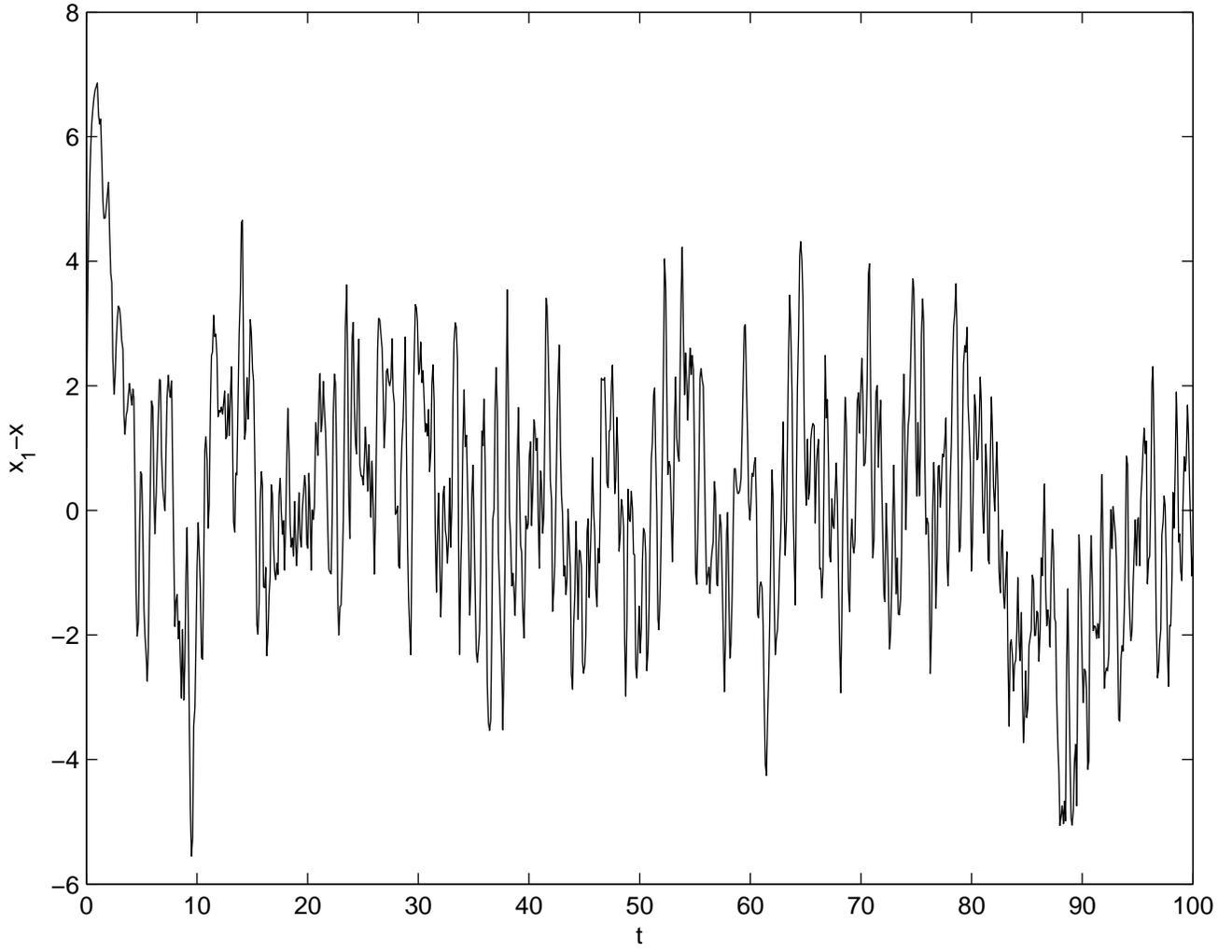
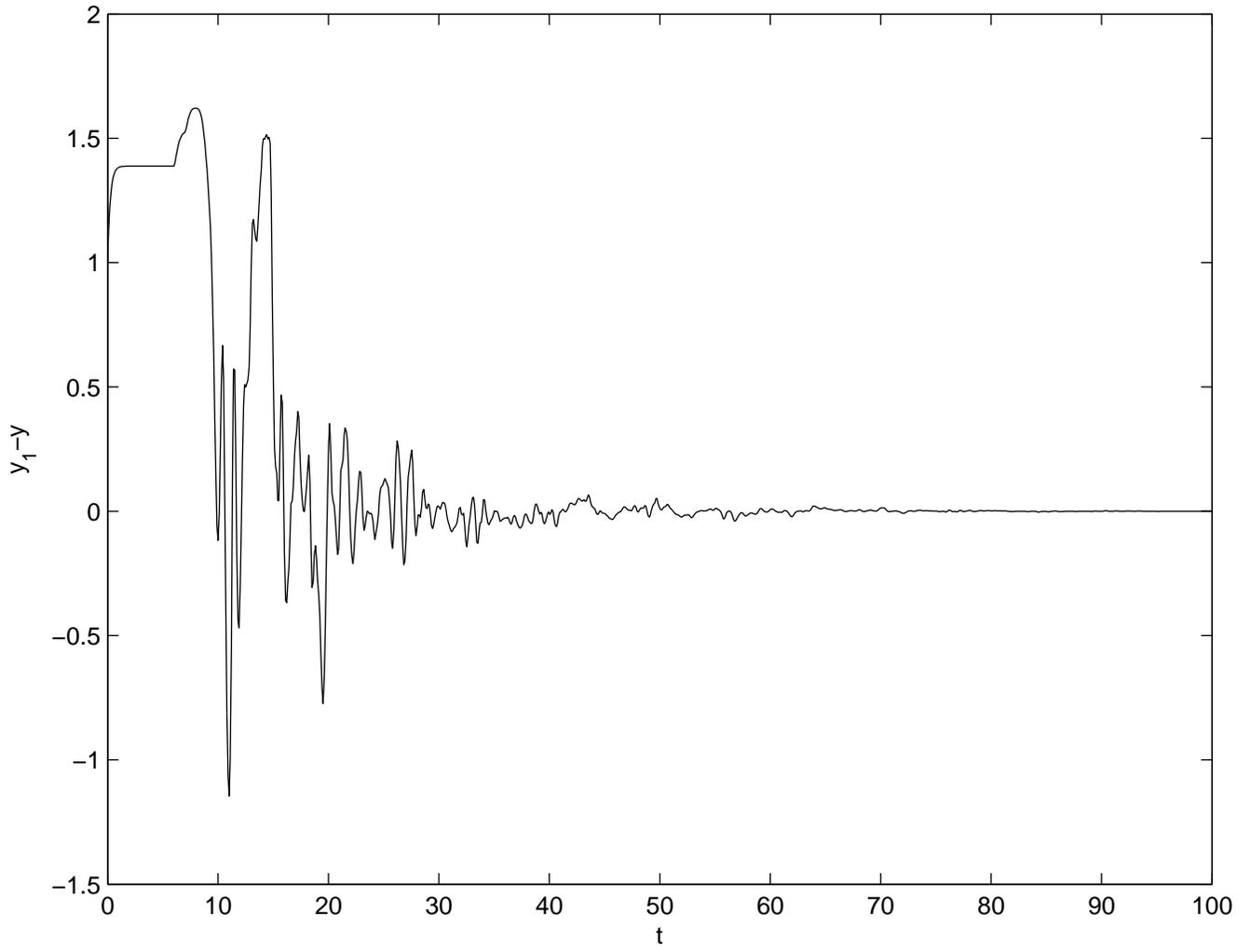
\noindent FIG.6. Numerical simulation of mutually coupled non-identical electro-optical laser systems, Eqs.(15-16): Partial generalized synchronization between $x$ and $y$ for $K_{1}=K_{2}=3.2$(other parameters as in Fig.5);(a)correlation plot between $x$ and $y;$(b)error $x_{1}-x$ dynamics;(c)error $y_{1}-y$ dynamics. C is the correlation coefficient between $x$ and $y.$ Dimensionless units.\\
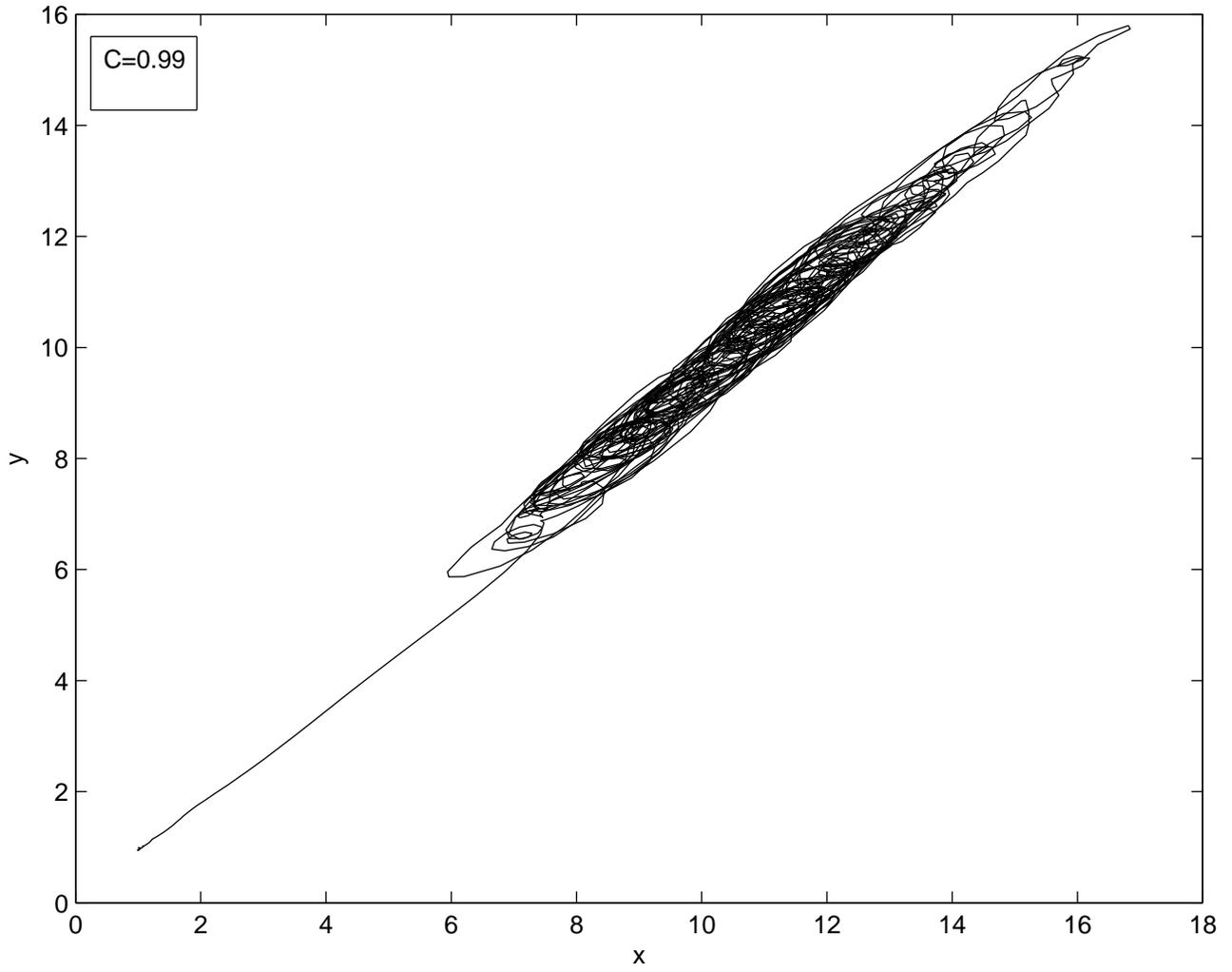
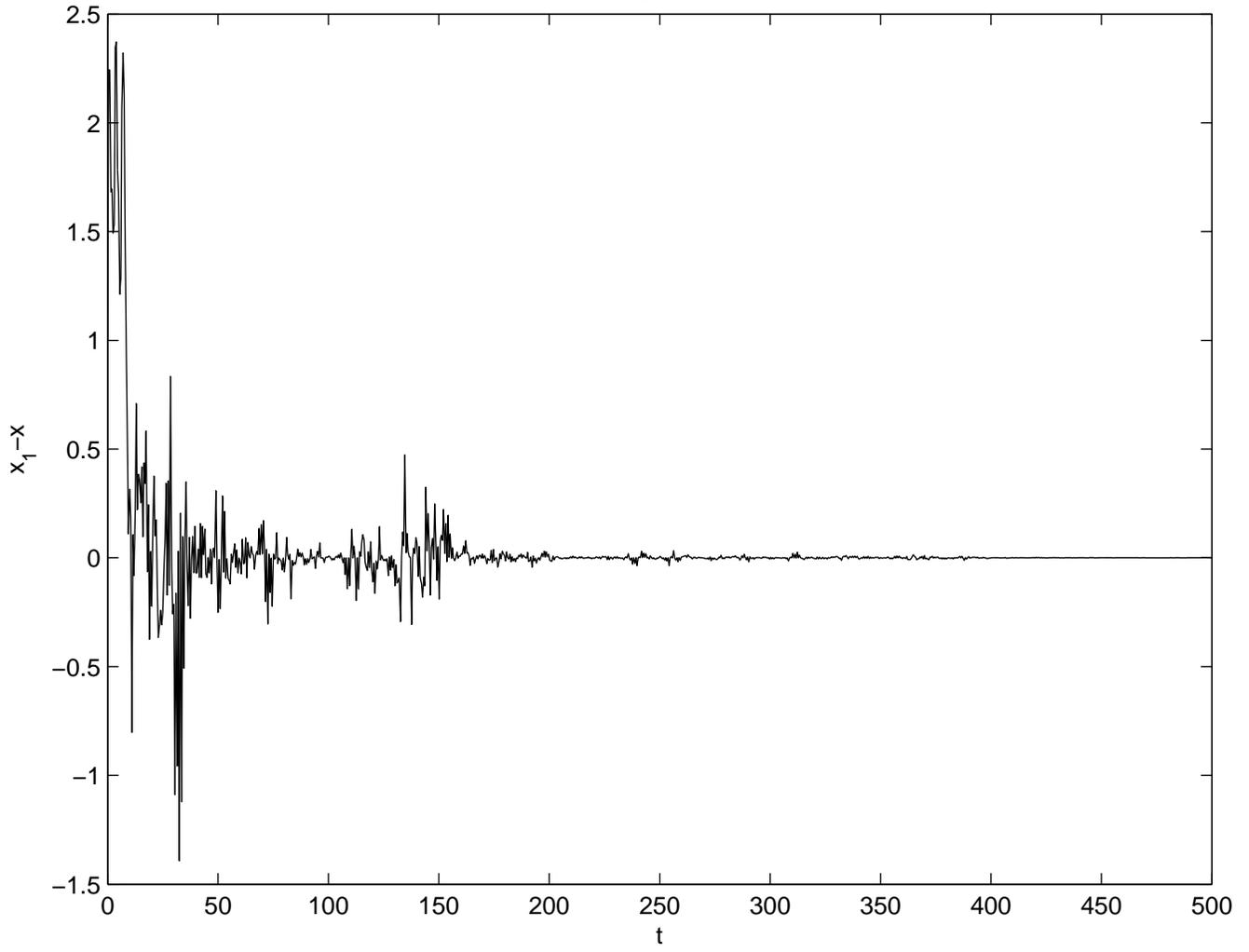
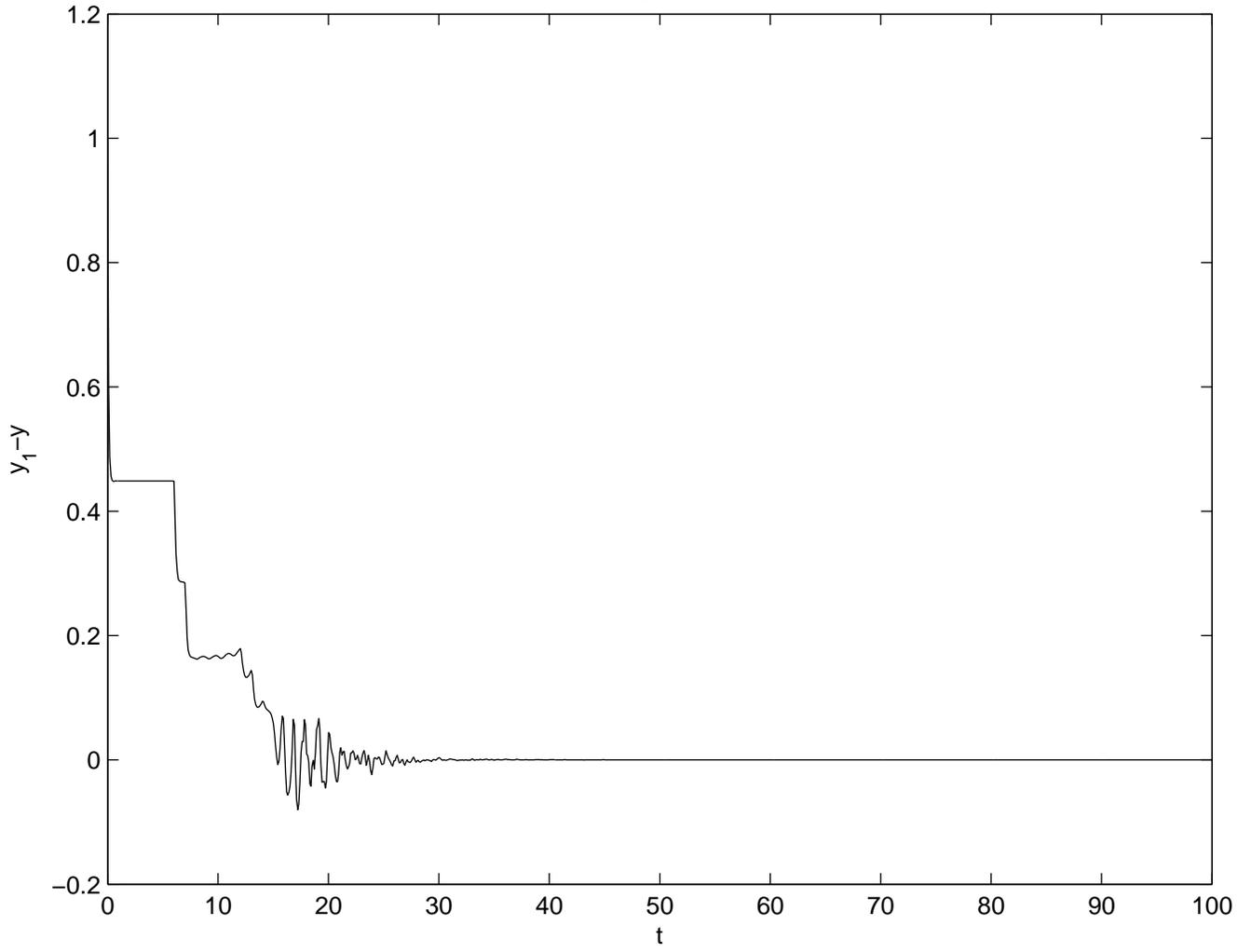
\noindent FIG.7. Numerical simulation of mutually coupled non-identical electro-optical laser systems, Eqs.(15-16): Global generalized synchronization between $x$ and $y$ for $K_{1}=K_{2}=12$(other parameters as in Fig.5) (a)correlation plot between $x$ and $y$(b)error $x_{1}-x$ dynamics;(c)error $y_{1}-y$ dynamics. C is the correlation coefficient between $x$ and $y.$ Dimensionless units.\\
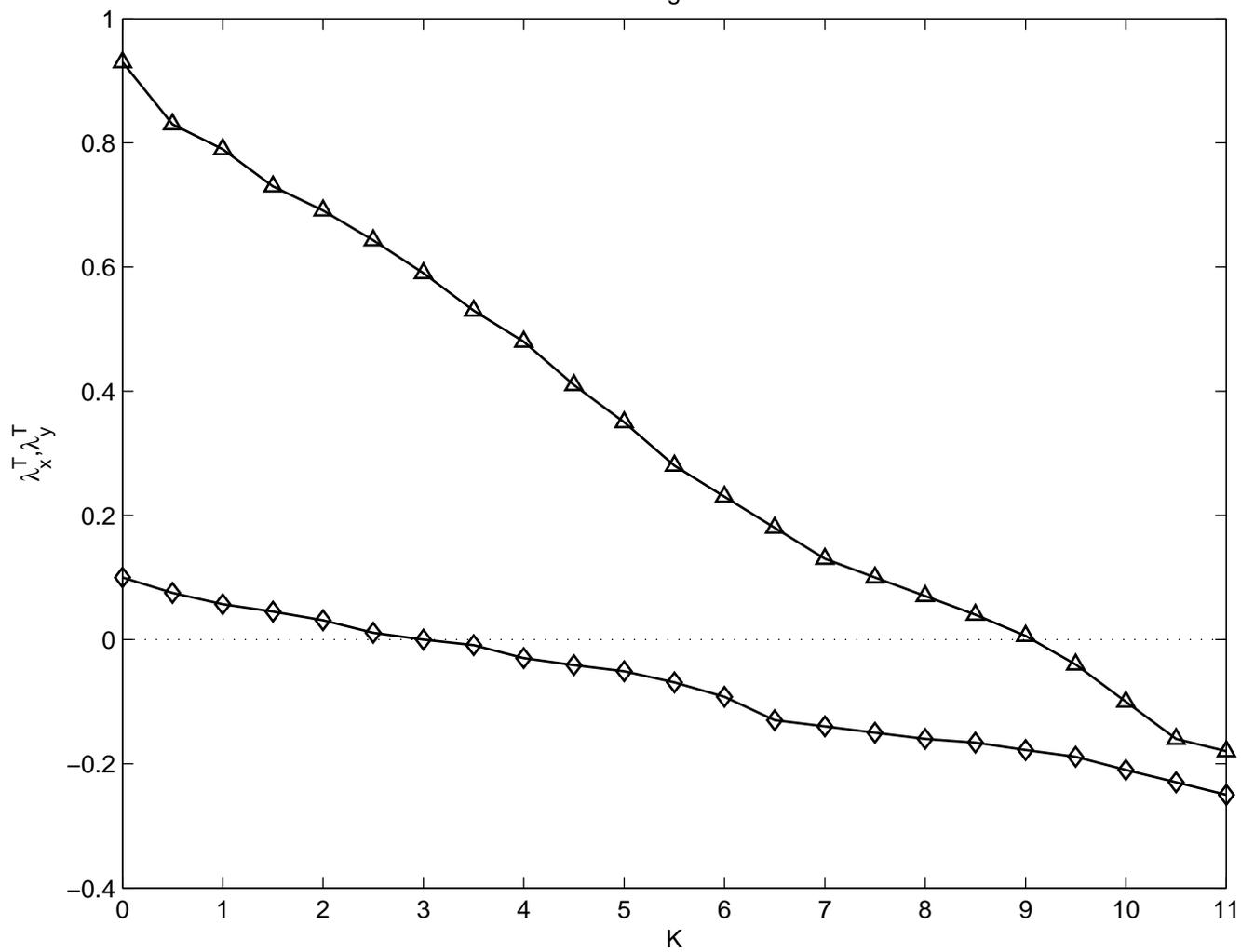
\noindent FIG.8. Numerical simulation of Eqs.(15-16), auxiliary systems' error dynamics $y_{1}-y,$ and  $x_{1}-x$  for $\Phi_{0}=\pi/4$,$m_{1}=20,
m_{2}=15, m_{3}=4, m_{4}=3,\tau_{1}=2, \tau_{2}=1,\tau_{3}=6,\tau_{4}=7.x$ laser system is more chaotic than $y$ laser: The maximum conditional Lyapunov exponents $\lambda_{x}^{T}$ ($\triangle $) and $\lambda_{y}^{T}$ ($\diamondsuit $) against the coupling strength  K for the auxiliary systems $x_{1}$ and $x$ driven by $y$ and $y_{1}$ and $y$ systems driven by $x,$ respectively. Dimensionless units.\\
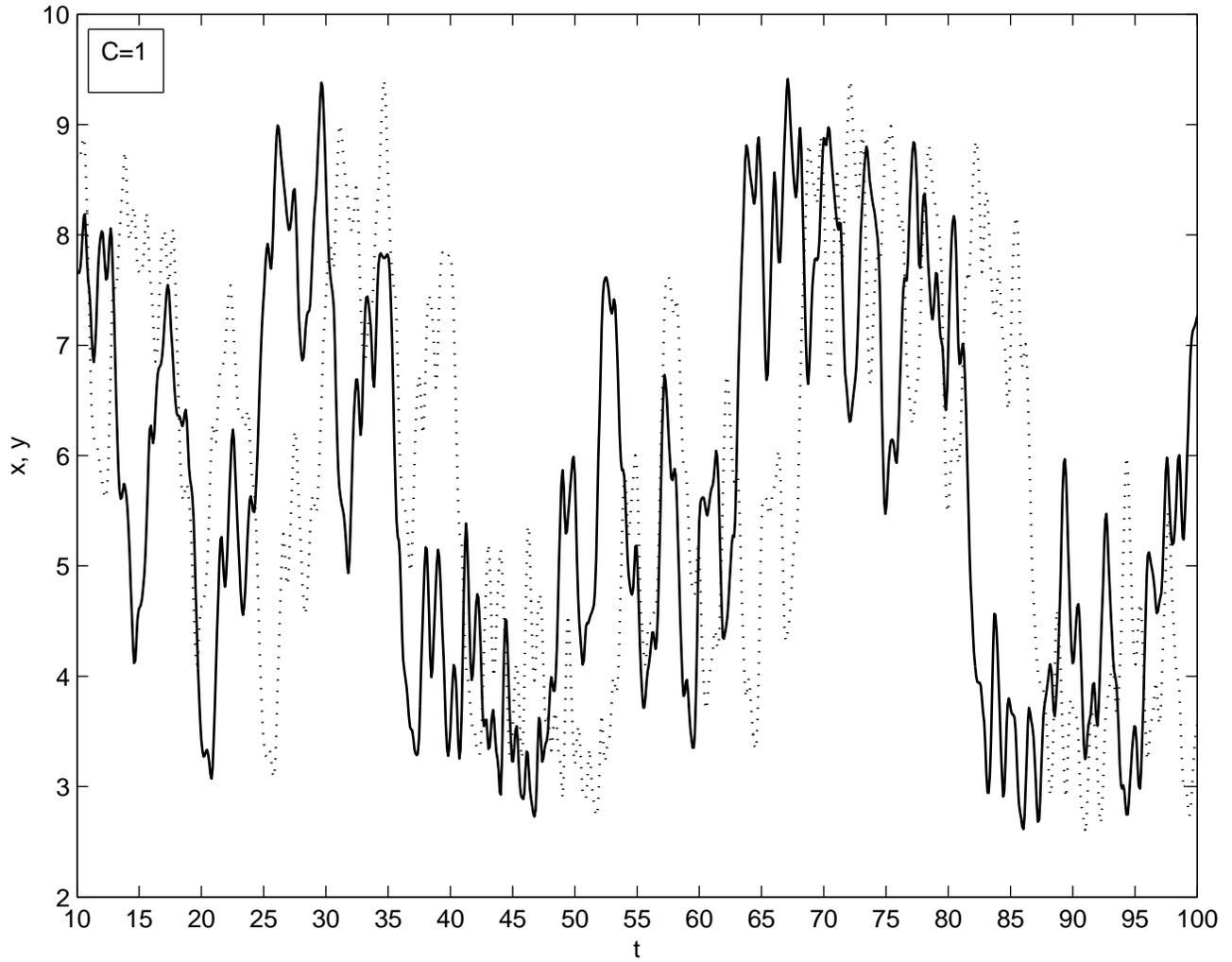
\noindent FIG.9. Numerical simulation of uni-directionally coupled non-identical electro-optical laser systems, Eqs.(21-22): Retarded synchronization with the coupling delay time between $x$(solid line) and $y$(dotted line) for $\alpha_{1}=1, \alpha_{2}=18, \Phi_{0}=\pi/4$, $m_{1}=m_{3}=7, m_{2}=m_{4}=5,K=17,\tau_{1}=3,\tau_{2}=2, \tau_{3}=5;$C is the correlation coefficient between $x_{\tau_{3}}$ and $y.$ Dimensionless units. \\

\end{document}